\documentclass[twocolumn,showpacs,preprintnumbers]{revtex4}
\usepackage{amssymb}
\usepackage{amsmath}
\usepackage{graphicx}
\usepackage{dcolumn}
\usepackage{bm}
\usepackage{color}
\usepackage{soul}

\soulregister\cite7
\soulregister\ref7

\begin{document}

\title{Double degeneracy associated with hidden symmetries in the asymmetric
two-photon Rabi model}
\author{You-Fei Xie and Qing-Hu Chen$^{*}$}

\address{
 Zhejiang Province Key Laboratory of Quantum Technology and Device,
 Department of Physics, Zhejiang University, Hangzhou 310027, China }
\date{\today }

\begin{abstract}
In this paper, we uncover the elusive level crossings in a subspace of the
asymmetric two-photon quantum Rabi model (tpQRM) when the bias parameter of
qubit is an even multiple of the renormalized cavity frequency. Due to the
absence of any explicit symmetry in the subspace, this double degeneracy
implies the existence of the hidden symmetry. The non-degenerate exceptional
points are also given completely. It is found that the number of the doubly
degenerate crossing points in the asymmetric tpQRM is comparable to that in
asymmetric one-photon QRM in terms of the same order of the constrained
conditions. The bias parameter required for occurrence of level crossings in
the asymmetric tpQRM is characteristically different from that at a multiple
of the cavity frequency in the asymmetric one-photon QRM, suggesting the
different hidden symmetries in the two asymmetric QRMs.
\end{abstract}

\maketitle

\section{ Introduction}

The simplest interaction between a two-level system (qubit) and a single
mode bosonic cavity (oscillator) was described by the quantum Rabi model
(QRM) \cite{Rabi,Braak2}, which is thus a fundamental textbook model in
quantum optics \cite{book}. It has been demonstrated in many advanced solid
devices, such as circuit quantum electrodynamics (QED) system \cite%
{Niemczyk,Forn2}, trapped ions \cite{Wineland}, and quantum dots \cite%
{Hennessy} from weak coupling to the ultra-strong coupling, even deep strong
coupling between the artificial atom and resonators \cite%
{Forn1,Yoshihara,Forn3}.

In contrast to the conventional cavity QED system, the artificial qubit
appears in modern solid devices usually contains both the splitting $\Delta $
and the bias $\epsilon $ between the two qubit states, thus the so-called
asymmetric QRM is ubiquitous. Driven by the proposals and experimental
realizations of the various QRMs model, the asymmetric two-photon QRM
(tpQRM) are also realized or stimulated to explore new quantum effects \cite%
{Bertet,Felicetti,tiefu}. The typical two asymmetric QRMs can be generally
written in a unified way as
\begin{equation}
H_{p}^{\epsilon }=\frac{\Delta }{2}\sigma _{z}+\frac{\epsilon }{2}\sigma
_{x}+\omega a^{\dag }a+g\left[ \left( a^{\dag }\right) ^{p}+a^{p}\right]
\sigma _{x},  \label{Hamiltonian}
\end{equation}%
where the first two terms fully describe a qubit with the energy splitting $%
\Delta $ and the bias $\epsilon $, $\sigma _{x,z}$ are the Pauli matrices, $%
a^{\dag }$ and $a$ are the creation and annihilation operators with the
cavity frequency $\omega $, and $g$ is the qubit-cavity coupling strength. $%
p=1,2$ denote the one-photon and two-photon QRMs, respectively. In the
superconducting flux qubit \cite{Forn1,Yoshihara}, $\Delta $ is the tunnel
coupling between the two persistent current states, $\epsilon =2I_{p}\left(
\Phi -\Phi _{0}/2\right) $ with $I_{p}$ the persistent current in the qubit
loop, $\Phi $ the an externally applied magnetic flux, and $\Phi _{0}$ the
flux quantum. The flux qubit is usually manipulated by the external magnetic
flux and the persistent currents.

For the symmetric case ($\epsilon =0$), the one-photon QRM possesses $%
\mathbb{Z}_{2}$-symmetry (parity), i.e. $[H_{1}^{0},\hat{P}_{1}]=0$ with the
parity operator $\hat{P}_{1}=\sigma _{z}\exp (i\pi a^{\dag }a),$ whose
eigenvalues are ${\pm 1}$, while the tpQRM has $\mathbb{Z}_{4}$-symmetry,
i.e. $[H_{2}^{0},\hat{P}_{2}]=0$ with the parity operator $\hat{P}%
_{2}=\sigma _{z}\exp (i\pi a^{\dag }a/2)$, whose eigenvalues are the quartic
roots of unity ${\pm 1,\pm i}$~\cite{duan2016,Felicetti15}. Hence the whole
Hilbert space separates therefore into two and four infinite-dimensional
subspaces in one-photon QRM and the tpQRM, respectively.

An analytical exact solution of the one-photon QRM has been found by Braak
in the Bargmann space representation \cite{Braak}. It was quickly reproduced
in the more familiar Hilbert space using the Bogoliubov operator approach
(BOA) by Chen \textsl{et al.} \cite{Chen2012}. Moreover, the BOA can be
easily extended to the tpQRM, and solutions in terms of a G-function, which
shares the common pole structure with Braak's G-function for the one-photon
QRM, are also found. It was soon realized that the G-function can be
constructed in terms of the mathematically well-defined Heun confluent
function \cite{Zhong}. These studies have stimulated extensive interests in
various QRMs ~\cite%
{Zhangyy,wanghui,Maciejewski21,duanEPL,luo2,bat,Zhiguo,Cong19,Xie2020}. For
more theoretical details in this field, one may refer to recent review
articles~ \cite{reviewJPA,Boite,Choi}.

The presence of the qubit bias term $\frac{\epsilon }{2}\sigma _{x}$ breaks $%
\mathbb{Z}_{2}$-symmetry of the QRM, so no any obvious symmetry remains in
the asymmetric QRM \cite{Zhong,bat,Wakayama}, while in the tpQRM, it reduces
the original $\mathbb{Z}_{4}$-symmetry to the $\mathbb{Z}_{2}$-symmetry. In
the asymmetric tpQRM, the $\mathbb{Z}_{2}$-symmetry corresponding to the
parity operator $\hat{P}_{2p}=\exp (i\pi a^{\dag }a)$ only acts in the
bosonic Hilbert spaces, the whole Hilbert space then only divides into two
invariant subspaces: even and odd number Fock states, which can be still
labeled by the Bargmann index $q=1/4$ and $3/4$ \cite{duan2016}.

Level crossing is very helpful to identify the symmetry in quantum systems.
The quasi-exact energies in the symmetric QRMs, also called Juddian
solutions \cite{Judd}, have been found 20 years ago ~\cite{Emary}. The
Juddian solutions are corresponding to the doubly degenerate states, and can
be constructed with the terminated polynomials. These quasi-exact energies
now can also be easily derived with the help of the pole structure of the $G$%
-function in both the one-photon \cite{Braak} and the two-photon \cite%
{Chen2012} QRMs. Surprisingly, the level crossing even exists without $%
\mathbb{Z}_{2}$-symmetry in the asymmetric one-photon QRM, when $\epsilon $
is a multiple of the cavity frequency $\omega $ \cite{Zhong}. In these
special cases, the hidden symmetry beyond any known symmetry is recently
discussed based on the numerical calculation on the energy eigenstates \cite%
{ash2020} and conserved operators~\cite{man,rey}.

For the symmetric tpQRM, the standard Juddian solutions are level crossings
within the same $q$ subspace. The second type of level crossings of the
eigenstates in different $q$ subspaces~\cite{Emary} was also found recently
\cite{Andrzej,xie2020}. In the asymmetric tpQRM, since $\mathbb{Z}_{4}$%
-symmetry reduces to $\mathbb{Z}_{2}$-symmetry, the level crossings within
the same $q$ subspace would generally disappear, while the second type of
the level crossings in the different $q$ subspaces remains robust due to the
remaining $\mathbb{Z}_{2}$-symmetry. Contrary to the one-photon asymmetric
QRM \cite{Zhong}, the level crossing within the same $q$ subspace in the
asymmetric tpQRM is elusive, and has not been observed to date. In this
work, we will uncover such a kind of level crossings irrelevant to any
explicit symmetry.

The paper is structured as follows: In Sec. II, we briefly review the
solutions to the asymmetric one-photon QRM in the framework of BOA approach,
and corroborate the previous observed doubly degenerate states in BOA frame.
We extend the BOA to study the asymmetric tpQRM, and derive the analytical
exact solutions in Sec. III. In Sec. IV, we discuss the non-degenerate
exceptional solutions for the asymmetric tpQRM. We demonstrate the level
crossings within the same $q$ subspace of the asymmetric tpQRM in Sec. V. 
The characteristics of level crossings in the two asymmetric  QRMs is discussed in Sec. VI. The last section contains some
concluding remarks. Appendix A confirms the conjecture that the two
vanishing coefficients give the same solutions in both asymmetric QRMs both
analytically in the  low order of and numerically in the large order of the
constrained conditions for the level crossings.

\section{Asymmetric quantum Rabi Model in BOA}

For the asymmetric one-photon QRM, when the bias parameter $\epsilon $ is a
multiple of the cavity frequency, the level crossings appear again in the
spectra even without any explicit known symmetry in the system~\cite%
{Zhong,bat}. {\ It should be noted that here $\epsilon $ in accord with the
standard qubit Hamiltonian ~\cite{Forn1,Yoshihara,ash2020,lizimin1} is twice
of that used in \cite{Braak,Zhong,bat}. }

In this section, we revisit the asymmetric one-photon QRM by BOA. We first
briefly review the solutions in the BOA framework ~\cite{Chen2012}, then we
can describe the level crossings in the BOA alternatively, which is
essentially equivalent to the Bargmann space approach. Furthermore, by BOA,
we can obtain all the non-degenerate exceptional points in a more concise
and complete way. Most importantly, this scheme can be easily extended to
the asymmetric tpQRM in the next sections.

\subsection{Solutions in BOA}

By two Bogoliubov transformations
\begin{equation}
A=a+g/\omega ,B=a-g/\omega ,
\end{equation}%
the wavefunction can be expressed as the series expansions in terms of $A$
operator
\begin{equation}
\left\vert A\right\rangle =\left( \
\begin{array}{l}
\sum_{n=0}^{\infty }\sqrt{n!}e_{n}\left\vert n\right\rangle _{A} \\
\sum_{n=0}^{\infty }\sqrt{n!}f_{n}\left\vert n\right\rangle _{A}%
\end{array}%
\right) ,  \label{wave1}
\end{equation}%
where $e_{n}\;$and $f_{n}$ are the expansion coefficients, and also in terms
of $B$ operator
\begin{equation}
\left\vert B\right\rangle =\left( \
\begin{array}{l}
\sum_{n=0}^{\infty }\left( -1\right) ^{n}\sqrt{n!}c_{n}\left\vert
n\right\rangle _{B} \\
\sum_{n=0}^{\infty }\left( -1\right) ^{n}\sqrt{n!}d_{n}\left\vert
n\right\rangle _{B}%
\end{array}%
\right) ,  \label{wave2}
\end{equation}%
with two coefficients $c_{n}\;$and $d_{n}$. $\left\vert n\right\rangle
_{A}\; $\ and $\left\vert n\right\rangle _{B}$ are called extended coherent
states ~\cite{chenqh}.

By the Schr$\overset{..}{o}$dinger equation, we get the linear relation for
two coefficients $e_{m}\;$and $f_{m}\;$ with the same index $m$ as ~\cite%
{Chen2012}
\begin{equation}
e_{m}=\frac{\Delta }{2\left( m\omega -g^{2}/\omega +\frac{\epsilon }{2}%
-E\right) }f_{m},  \label{cor1}
\end{equation}%
and the coefficient $f_{m}$ can be defined recursively,
\begin{widetext}
\begin{equation}
\left( m+1\right) f_{m+1}=\frac{1}{2g}\left( m\omega +3g^{2}/\omega -\frac{%
\epsilon }{2}-E-\frac{\Delta ^{2}}{4\left( m\omega -g^{2}/\omega +\frac{%
\epsilon }{2}-E\right) }\right) f_{m}-f_{m-1},  \label{fm}
\end{equation}%
\end{widetext}
with\ $f_{0}=1$. Similarly, the two coefficients $c_{m}\;$and $d_{m}$
satisfy
\begin{equation}
d_{m}=\frac{\Delta }{2\left( m\omega -g^{2}/\omega -\frac{\epsilon }{2}%
-E\right) }c_{m},  \label{cor2}
\end{equation}%
and the recursive relation is given by
\begin{widetext}
\begin{equation}
\left( m+1\right) c_{m+1}=\frac{1}{2g}\left( m\omega +3g^{2}/\omega +\frac{%
\epsilon }{2}-E-\frac{\Delta ^{2}}{4\left( m\omega -g^{2}/\omega -\frac{%
\epsilon }{2}-E\right) }\right) c_{m}-c_{m-1},  \label{cm}
\end{equation}
\end{widetext}
with $c_{0}=1.$

If both wavefunctions (\ref{wave1}) and (\ref{wave2}) are the true
eigenfunction for a non-degenerate eigenstate with eigenvalue $E$, they
should be in principle only different by a complex constant $z$, i.e. $%
\left\vert A\right\rangle =z\left\vert B\right\rangle $. Projecting both
sides onto the original vacuum state $\left\vert 0\right\rangle $, using $%
\sqrt{n!}{\langle }0|n{\rangle }_{A}=(-1)^{n}\sqrt{n!}{\langle }0|n{\rangle }%
_{B}=e^{-\left( \frac{g}{\omega }\right) ^{2}/2}\left( \frac{g}{\omega }%
\right) ^{n}$ and eliminating the ratio constant $z$ gives
\begin{equation}
\sum_{n=0}^{\infty }e_{n}\left( \frac{g}{\omega }\right)
^{n}\sum_{n=0}^{\infty }d_{n}\left( \frac{g}{\omega }\right)
^{n}=\sum_{n=0}^{\infty }f_{n}\left( \frac{g}{\omega }\right)
^{n}\sum_{n=0}^{\infty }c_{n}\left( \frac{g}{\omega }\right) ^{n},
\label{prop}
\end{equation}%
with the help of Eqs. (\ref{cor1}) and (\ref{cor2}), one arrives at
one-photon G-function
\begin{eqnarray}
G_{1p} &=&\left( \frac{\Delta }{2}\right) ^{2}\left[ \sum_{n=0}^{\infty }%
\frac{f_{n}}{n\omega -g^{2}/\omega +\frac{\epsilon }{2}-E}\left( \frac{g}{%
\omega }\right) ^{n}\right]  \notag \\
&&\times \left[ \sum_{n=0}^{\infty }\frac{c_{n}}{n\omega -g^{2}/\omega -%
\frac{\epsilon }{2}-E}\left( \frac{g}{\omega }\right) ^{n}\right]  \notag \\
&&-\sum_{n=0}^{\infty }f_{n}\left( \frac{g}{\omega }\right)
^{n}\sum_{n=0}^{\infty }c_{n}\left( \frac{g}{\omega }\right) ^{n}.
\label{G_1p}
\end{eqnarray}%
This G-function was first derived by Braak \cite{Braak} using Bargmann space
approach, and later reproduced by Chen et al. \cite{Chen2012}. We then
discuss the level crossing of this asymmetric QRM in terms of BOA framework
described above.

\subsection{Doubly degenerate states}

The two types of pole energies appear in the one-photon G-function (\ref%
{G_1p}) as
\begin{eqnarray}
E_{N}^{A} &=&N\omega -g^{2}/\omega +\frac{\epsilon }{2},N=0,1,2,...
\label{pole_1pA} \\
E_{M}^{B} &=&M\omega -g^{2}/\omega -\frac{\epsilon }{2},M=0,1,2,...
\label{pole_1pB}
\end{eqnarray}%
They are labeled with the type-A and type-B pole energy, respectively. If
\begin{equation}
\epsilon =\left( M-N\right) \omega ,  \label{epsilon_p1}
\end{equation}%
these two pole energies are the same
\begin{equation}
E_{N}^{A}(g)=E_{M}^{B}(g)=\frac{1}{2}\left( M+N\right) \omega -g^{2}/\omega .
\label{poleline}
\end{equation}%
Note that $\epsilon $ should be a multiple of the cavity frequency $\omega$
under the condition (\ref{epsilon_p1}). In this paper, we only consider $M>N$%
, so that $\epsilon $ is positive. For the case of $M<N$, the extension is
achieved straightforwardly by changing $\epsilon $ into $-\epsilon $ and
interchanging $M$ and $N$.

From Eqs. (\ref{cor1}) [(\ref{cor2})], one immediately notes that the
coefficient $e_{N}$ ($d_{M}$) would diverge at the same pole energy (\ref%
{poleline}). It does not make sense if some coefficients in the series
expansion of a wavefunction really become infinity. A normalizable
wavefunction should consist of the global property, i.e. the finite inner
product, so the series expansion coefficients in the wavefunction (\ref%
{wave1}) and (\ref{wave2}) should be analytic and vanish as or before $%
n\rightarrow \infty $.

To achieve a physics state, at the pole energy (\ref{poleline}), the
numerator of right-hand-side of Eq. (\ref{cor1}) [(\ref{cor2})] should also
vanish, so that $e_{N}$ ($d_{M}$) remains finite, which result in
\begin{equation}
f_{N}(M,g)=0;c_{M}(N,g)=0.  \label{location}
\end{equation}%
Note that $f_{N}$ and $c_{M}$ can be obtained using the following
three-terms recurrence relation from (\ref{fm}) and (\ref{cm}) with energy (%
\ref{poleline}), respectively
\begin{eqnarray}
\left( n+1\right) f_{n+1}&=&\frac{1}{2g}\left[ 4g^{2}/\omega +\left(
n-M\right) \omega -\frac{\Delta ^{2}}{4\omega \left( n-N\right) }\right]f_{n}
\notag \\
&&-f_{n-1},  \label{r_fN}
\end{eqnarray}
\begin{eqnarray}
\left( n+1\right) c_{n+1}&=&\frac{1}{2g}\left[ 4g^{2}/\omega +\left(
n-N\right) \omega -\frac{\Delta ^{2}}{4\omega \left( n-M\right) }\right]c_{n}
\notag \\
&&-c_{n-1}.  \label{r_fpm}
\end{eqnarray}
If $\Delta ,M,N$ are given, two equations in (\ref{location}) would provide
the coupling strength $g$ in the energy spectra where the energy levels
intersect with the same pole line described by Eq. (\ref{poleline}).

A mathematical proof to the conjecture that $f_{N}(M,g)=0$ and $c_{M}(N,g)=0$
could give the same real and positive solutions for the coupling strength $g$
was given in \cite{Wakayama}. Li and Batchelor \cite{bat} have analyzed the
relation between the number of the exceptional points and the model
parameters ($\Delta $ and $\epsilon $), and numerically found that the
number of positive roots from these two equations are the same for integer $%
\epsilon/\omega $. {But we confine us here to a closed-form proof for small
values of $N$ and $M$, and numerically confirmation for large $N$ and $M$.
In the asymmetric tpQRM, similar constrained condition will be derived and
we will also do the similar things because a similar conjecture will be
proposed but cannot be proven at the present stage. }

To this end, we present our discussions only in terms of the fixed integers $%
N$ and $M$. In this case, $\epsilon $ is a multiple of the cavity frequency
is known immediately, and the remaining task is to show the same crossing
points by two equations in Eq. (\ref{location}), which is illustrated in
Appendix A1.

\begin{figure}[tbp]
\includegraphics[width=\linewidth]{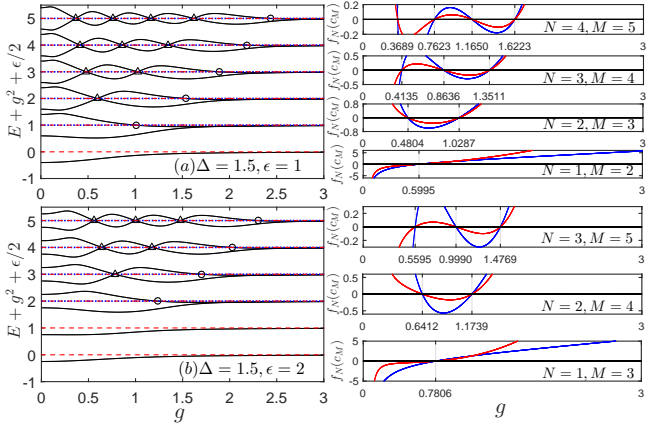}
\caption{ (Color online) Energy spectrum $E+g^{2}+\protect\epsilon /2$ for $%
\protect\omega=1, \Delta =1.5,\protect\epsilon =1$ (a) and $\protect\epsilon %
=2$ (b) in left panels. The horizontal blue dotted lines correspond to the
pole energy ones for $E_{N}^{A}$ and the red dashed lines to $E_{M}^{B}$.
Only the overlapped pole lines with $N>0$ allow for the true level
crossings. The triangles denotes the doubly degenerate crossing points.
Circles indicate the non-degenerate exceptional solutions by Eq. (\protect
\ref{non_M}). $f_{N}(M,g)$ (blue) and $c_{M}(N,g)$ (red) curves are shown in
the right panels. The zeros are exactly corresponding to the triangles in
the left spectrum.}
\label{1pcrossD15}
\end{figure}

\begin{figure}[tbp]
\includegraphics[width=\linewidth]{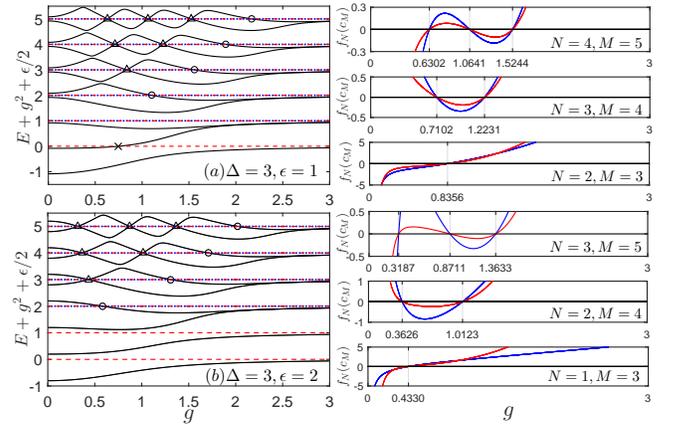}
\caption{ (Color online) Notations are the same as those in Fig. \protect\ref%
{1pcrossD15} except for $\Delta =3$. Note that the doubly degenerate
crossing point is absent in the $N=1, M=2$ overlapped line in (a) due to $%
\Delta >2\protect\sqrt{2}$ here. }
\label{1pcrossD30}
\end{figure}

{\ In the left panels of Figs. \ref{1pcrossD15} and \ref{1pcrossD30}, we
present the energy spectrum }${E+g}^{2}/\omega +\frac{\epsilon }{2}${\ } {%
for }$\epsilon =1$, $2$ at $\Delta =1.5$ {\ }and $\Delta =3$ with $\omega =1$%
, respectively. The dotted and dashed horizontal lines denote different
types of pole lines. Obviously, if the two types of pole lines cannot
coincide, the level crossings cannot happen. $f_{N}(M,g)$ and $c_{M}(N,g)$
curves are plotted in the right panels. By Eq. (\ref{gN1}), one finds $%
g=\allowbreak 0.5995$ for $N=1,M=2\ $at $\Delta =1.5$, consistent with the
spectra in Fig. \ref{1pcrossD15} (a). For $\Delta =3$, see Eq. (\ref{1pN1M2}%
), no real positive solution can be found in this case, so level crossings
cannot occur in the overlapped line with $N=1$ in Fig. \ref{1pcrossD30} (a).

At $\Delta =1.5$, we can obtain two solutions for $g$ as $0.4804$, $1.0287$
for $N=2$ and $M=3$ by Eq. (\ref{N2M3}), agreeing well with two crossing
points in the second type-A pole line with $N=2$ and $M=3$ shown in Fig. \ref%
{1pcrossD15} (a). For $\Delta =3,$ we only find one real positive $%
g=\allowbreak 0.8356$, consistent with the spectra in Fig. \ref{1pcrossD30}%
(a).

Associated with the overlapped $N=1$ type-A and $M = 3$ pole lines, one can
obtain $g=0.7806$ for $\Delta =1.5$, and $g=0.4330$ for $\Delta =3$ by Eq. (%
\ref{N1M3}), consistent with the crossing points in the calculated spectrum
in Fig. \ref{1pcrossD15} (a) and Fig. \ref{1pcrossD30} (a).

For large value of $M$ and $N$, as shown in the right panels of Figs. \ref%
{1pcrossD15}\ and \ref{1pcrossD30}, both $f_{N}(M,g)$ and $c_{M}(N,g)$
curves provide the same zeros for all cases.

Now we will further demonstrate explicitly that any {crossing point found
above is corresponding to a doubly degenerate state in the BOA framework. At
the crossing point, looking at (\ref{cor1}), since both the numerator }$%
f_{N}(g)$ and denominator vanish, $e_{N}$ would be arbitrary. If we set
\begin{equation}
e_{N}=-\frac{4g}{\Delta }f_{N-1},  \label{special}
\end{equation}%
from Eq. (\ref{fm}) we know $f_{N+1}=0$, further $e_{N+1}=0$, and all
coefficients $f_{k}$ and $e_{k}$ for $k>N+1$ vanish. So the infinite series
expansion in the wavefunction (\ref{wave2}) terminates with finite $N$ as
\begin{equation}
\left\vert A\right\rangle _{N}=\left( \
\begin{array}{l}
\sum_{n=0}^{N}\sqrt{n!}e_{n}\left\vert n\right\rangle _{A} \\
\sum_{n=0}^{N-1}\sqrt{n!}f_{n}\left\vert n\right\rangle _{A}%
\end{array}%
\right) .  \label{wterm1}
\end{equation}%
Similarly, the infinite series expansion in the wavefunction (\ref{wave2})
terminates with finite $M$ as
\begin{equation}
\left\vert B\right\rangle _{M}=\left( \
\begin{array}{l}
\sum_{n=0}^{M-1}\left( -1\right) ^{n}\sqrt{n!}c_{n}\left\vert n\right\rangle
_{B} \\
\sum_{n=0}^{M}\left( -1\right) ^{n}\sqrt{n!}d_{n}\left\vert n\right\rangle
_{B}%
\end{array}%
\right) ,  \label{wterm2}
\end{equation}%
where
\begin{equation*}
d_{M}=-\frac{4g}{\Delta }c_{M-1}.
\end{equation*}%
Interestingly, both wavefunction terminates at finite terms. Because these
two wavefuntions are not obtained from the G-function based on the
proportionality (\ref{prop}), so they are different $\left\vert
A\right\rangle _{N}\neq \left\vert B\right\rangle _{M}$, leading to {doubly
degenerate states}. Since the degenerate eigenfunctions, $\left\vert
A\right\rangle _{N}$ and $\left\vert B\right\rangle _{M}$, are given as
finite polynomials in the extended coherent state basis $\{\left\vert
n\right\rangle _{A}\}$ and $\{\left\vert n\right\rangle _{B}\}$ (see also~%
\cite{Braak19}). These states are the quasi-exact solutions of the
asymmetric QRM.

At this stage, we can simply discuss the number of the doubly degenerate
crossing points associated with the given $N$ type-A pole line. $f_{N}(M>N,
x=4g^{2})$ derived by Eq. (\ref{r_fN}) is a polynomial with $N$ terms. Its
zero would generally give around $N$ roots, indicating that there are around
$N$ doubly degenerate crossing points along the $N$ type-A pole line in the
energy spectra. Note that for large $\Delta$, the number of the roots could
be slightly less than $N$, as shown in Fig. \ref{1pcrossD30}. For small $%
\Delta$, we can actually have just $N$ roots.

\subsection{Non-degenerate exceptional points}

The non-degenerate exceptional points can be generated if only one energy
level intersects with the energy pole line alone. In principle, all
non-degenerate states including non-degenerate exceptional ones can be
obtained by the G-function (\ref{G_1p}) because it is built based on the
proportionality (\ref{prop}), only excluding the degenerate states. These
states have been first analyzed for the symmetric QRM with the Bargmann
space technique in \cite{Maciejewski21} and later in \cite%
{Braak19,braak-fmi,xychen}. We believe that the BOA has advantages with
regard to the non-degenerate exceptional solutions, which cannot be found
with any ansatz.

Note from G-function (\ref{G_1p}) that, at the pole energy either (\ref%
{pole_1pA}) or (\ref{pole_1pB}), the denominator of the associated term
become zero, so this term would diverge and should be treated specially. For
a physics state, to avoid the divergence, the {numerator }$f_{N}(g)$ or $%
c_{M}(g)$ should also vanish. It is very important to see that {\ }$f_{N}(g)$
or $c_{M}(g)$ could vanish in two different ways. First, $f_{n}$ ($c_{m}$)
can be obtained by using the three-terms recurrence relation (\ref{fm}) [(%
\ref{cm})] from $f_{0}=1$ ($c_{0}=1$), and $f_{N}=0$ ($c_{M}=0$) until $n=N$
($m=M$). Second, one can set $f_{n\leqslant N}=0$ ($c_{m\leqslant M}=0$) and
$e_{n=N}=1$ ($d_{m=M}=1$) at the beginning directly and obtain all remaining
coefficients by the recurrence relation (\ref{fm})[(\ref{cm})]. This is to
say, we have two ways to overcome the divergence. In the infinite summation
where the diverging term is present, we may cut off all the terms either
after or before this diverging one. E.g. for the $N$th type-A pole line, we
may terminate the infinite summation at the diverging term following the
same idea outlined in the last section for the degenerate states. So the
first non-degenerate exceptional G-function can be written as
\begin{eqnarray}
G_{1p}^{non,1A} &=&\left[ \sum_{n=0}^{N-1}\frac{\Delta f_{n}}{2\omega \left(
n-N\right) }\left( \frac{g}{\omega }\right) ^{n}+e_{N}\left( \frac{g}{\omega
}\right) ^{N}\right]  \notag \\
&&\times \left[ \sum_{n=0}^{\infty }\frac{\Delta c_{n}}{2\left( n\omega
-g^{2}/\omega -\epsilon \right) }\left( \frac{g}{\omega }\right) ^{n}\right]
\notag \\
&&-\sum_{n=0}^{N-1}f_{n}\left( \frac{g}{\omega }\right)
^{n}\sum_{n=0}^{\infty }c_{n}\left( \frac{g}{\omega }\right) ^{n}=0,
\label{subset1}
\end{eqnarray}%
where $e_{N}$ \ is given by Eq. (\ref{special}). Note that the remaining
terms vanish because all coefficients become zero. We can also remove all
terms before the diverging term in the summation, and give the second
non-degenerate exceptional G-function as
\begin{eqnarray}
G_{1p}^{non,2A} &=&\left[ \left( \frac{g}{\omega }\right)
^{N}+\sum_{n=N+1}^{\infty }\frac{\Delta f_{n}}{2\omega \left( n-N\right) }%
\left( \frac{g}{\omega }\right) ^{n}\right]  \notag \\
&&\times \left[ \sum_{n=0}^{\infty }\frac{\Delta c_{n}}{2\left( n\omega
-g^{2}/\omega -\epsilon \right) }\left( \frac{g}{\omega }\right) ^{n}\right]
\notag \\
&&-\sum_{n=N+1}^{\infty }f_{n}\left( \frac{g}{\omega }\right)
^{n}\sum_{n=0}^{\infty }c_{n}\left( \frac{g}{\omega }\right) ^{n}=0,
\label{subset2}
\end{eqnarray}%
with the initial condition $e_{N}=1$. \ The non-degenerate exceptional
G-functions $G_{1p}^{non,1B}$ and $G_{1p}^{non,2B}$ associated with the
type-B pole line can be obtained similarly by modifying the other infinite
summation, which are not shown here.

Two non-degenerate exceptional G-functions (\ref{subset1}) and (\ref{subset2}%
) provide different exceptional solutions, which comprise the full
non-degenerate exceptional points associated with the Type-A pole lines.
Particularly, $f_{N}=0$ or $c_{M}=0$ is implied Eq. (\ref{subset1}) or $%
G_{1p}^{non,1B}=0$, thus can be also used to give the same non-degenerate
exceptional points in a simpler way. Just as pointed out in Ref \cite{bat},
for noninteger $\epsilon $, a subset of the non-degenerate exceptional
points associated with the pole lines can be given by the vanishing
coefficients $f_{m}$ or $c_{m}$, equivalently, using Eq. (\ref{subset1}) or $%
G_{1p}^{non,1B}=0$ here. However Eq. (\ref{subset1}) and $G_{1p}^{non,1B}=0$
fail at integer $\epsilon $ including $\epsilon =0$, because $f_{N}=0$ or $%
c_{M}=0$ actually results in the doubly degenerate states, which results in
nonzero G-function in this case.

Interestingly, for integer $\epsilon $, two types of pole line may merge
together. At the same pole energy (\ref{poleline}), the second $\ $%
non-degenerate exceptional G-function Eq. (\ref{subset2}) would be further
modified as
\begin{eqnarray}
G_{1p}^{non} &=&\left[ \left( \frac{g}{\omega }\right)
^{N}+\sum_{n=N+1}^{\infty }\frac{\Delta f_{n}}{2\omega \left( n-N\right) }%
\left( \frac{g}{\omega }\right) ^{n}\right]  \notag \\
&&\times \left[ \left( \frac{g}{\omega }\right) ^{M}+\sum_{n=M+1}^{\infty }%
\frac{\Delta c_{n}}{2\omega \left( n-M\right) }\left( \frac{g}{\omega }%
\right) ^{n}\right]  \notag \\
&&-\sum_{n=N+1}^{\infty }f_{n}\left( \frac{g}{\omega }\right)
^{n}\sum_{n=M+1}^{\infty }c_{n}\left( \frac{g}{\omega }\right) ^{n}=0,
\label{non_M}
\end{eqnarray}%
where $e_{n<N}=0,e_{N}=1,$ and $d_{n<M}=0,d_{M}=1$, the other coefficients
can still be obtained from the three-terms recurrence relations (\ref{fm})
and (\ref{cm}).

In the left panels of Figs. \ref{1pcrossD15} and \ref{1pcrossD30}, the
non-degenerate exceptional points are indicated by open circles and crosses
where the energy levels intersect with the pole lines alone. All the open
circles are given by zeros of the non-degenerate exceptional G-function (\ref%
{non_M}), while a cross in Fig. \ref{1pcrossD30} (a) is solved by G-function
$G_{1p}^{non,2B}$ associated with the type-B pole lines, c. f. Eq. (\ref%
{subset2}).

In the end of this section, we would like to point out that the previous
main results in the asymmetric QRM based on the Bargmann space approach, see
Ref ~\cite{bat} and reference therein, can be well described in the BOA
framework in a self-contained way. The asymmetric tpQRM has not been studied
in the literature, much less the level crossings irrelevant to the explicit
symmetry, to our knowledge. Note that the G-function by the direct
application of the Bargmann space approach to the tpQRM ~\cite{Trav} has no
pole structure, and thus could not give qualitative insight into the
behavior of the spectral collapse \cite{Felicetti15} and the level crossing.
As far as we know, the G-function with its pole structure for the tpQRM has
only been found using the BOA ~\cite{Chen2012, duan2016,Cui} and, in
particular, has so far not been derived using the Bargmann space method in
the literature. Therefore, it is perhaps irreplaceable, at the moment, to
employ the BOA to study the asymmetric tpQRM, which is the main topic of
this paper.

\section{Asymmetric two-photon Rabi Model and solutions using BOA}

For convenience, we rewrite the Hamiltonian $H_{2}^{\epsilon }$ on the $%
\sigma _{z}$ basis by rotating it around the $y$-axis with an angle $\pi /2$%
. The transformed Hamiltonian is given by the following matrix form
\begin{equation}
H_{2,r}^{\epsilon }=\left(
\begin{array}{cc}
\omega a^{\dag }a+g(a^{\dag 2}+a^{2})+\frac{\epsilon }{2} & -\frac{\Delta }{2%
} \\
-\frac{\Delta }{2} & \omega a^{\dag }a-g(a^{\dag 2}+a^{2})-\frac{\epsilon }{2%
}%
\end{array}%
\right) .  \label{H_0}
\end{equation}

The Hamiltonian above is connected with $su(1,~1)$ Lie algebra
\begin{equation}
K_{0}=\frac{1}{2}(a^{\dag }a+\frac{1}{2}),K_{+}=\frac{1}{2}a^{\dag 2},K_{-}=%
\frac{1}{2}a^{2},  \label{K_0}
\end{equation}%
which obey spin-like commutation relations $[K_{0},K_{\pm }]=\pm K_{\pm
},[K_{+},K_{-}]=-2K_{0}$. The quadratic invariant Casimir operator is given
by
\begin{equation*}
\mathcal{C}_{2}=K_{+}K_{-}+K_{0}\left( 1-K_{0}\right) .
\end{equation*}
Then we apply a squeezing operator $S_{1}=e^{\frac{r}{2}(a^{2}-a^{\dag 2})}$
to diagonalize the bosonic part of the above Hamiltonian and the parameter $%
r $ is to be fixed later. In terms of the $K_{0},K_{\pm }$, the transformed
Hamiltonian is derived as
\begin{equation}
H_{2,r}^{\epsilon }=\left(
\begin{array}{cc}
\beta \left( 2K_{0}\right) +\frac{\epsilon -\omega }{2} & -\frac{\Delta }{2}
\\
-\frac{\Delta }{2} & H_{22}%
\end{array}%
\right) ,  \label{H}
\end{equation}%
where $\beta =\omega \sqrt{1-4\left( \frac{g}{\omega }\right) ^{2}}<\omega $
can be termed as the renormalized cavity frequency owing to the fact that it
is just a g-dependent pre-factor of the free photon number operators $2K_0$,
and $\beta=\omega$ if $g=0$. It will be shown later that $\beta$ plays a key
role in two-photon QRM. The second diagonal element is
\begin{eqnarray*}
H_{22} &=&(2\omega \cosh 2r-4g\sinh 2r)K_{0} \\
&&+(\omega \sinh 2r-2g\cosh 2r)(K_{+}+K_{-})-\frac{\epsilon +\omega }{2},
\end{eqnarray*}%
and the squeezing parameter
\begin{equation}
r=\frac{1}{4}\ln \left( \frac{1-2g/\omega }{1+2g/\omega }\right) .  \label{r}
\end{equation}%
It is obvious that the coupling strength $g<\omega /2$ leads to a real
squeezing parameter.

Based on the squeezing transformation, we propose the corresponding
wavefunction as
\begin{equation}
\left\vert \Psi _{A}\right\rangle ^{q}=\binom{\sum_{m=0}^{\infty }\sqrt{%
[2(m+q-\frac{1}{4})]!}e_{m}^{(q)}\left\vert q,m\right\rangle _{A}}{%
\sum_{m=0}^{\infty }\sqrt{[2(m+q-\frac{1}{4})]!}f_{m}^{(q)}\left\vert
q,m\right\rangle _{A}},  \label{w1}
\end{equation}%
where the new basis $\left\vert q,m\right\rangle _{A}=S_{A}\left\vert
q,m\right\rangle $ with $\left\vert q,m\right\rangle $ is the Fock state.
The coefficients $e_{m}^{(q)}$ and $f_{m}^{(q)}$ are to be determined in the
following.

In the case of the Lie algebra considered here, $K_{0}\left\vert
q,0\right\rangle _{A}=q\left\vert q,0\right\rangle _{A}$ where $q=\frac{1}{4}
$ and $\frac{3}{4}$ divide the whole Hilbert space $\mathcal{H}$ into even
and odd sectors and label them, respectively. For the even subspace, $%
\mathcal{H}_{1/4}=\left\{ a^{\dag n}\left\vert 0\right\rangle
,n=0,2,4,...\right\} $, and for the odd subspace, $\mathcal{H}_{3/4}=\left\{
a^{\dag n}\left\vert 0\right\rangle, n=1,3,5,...\right\} $, corresponding to
even or odd Fock number basis. The Casimir element $\mathcal{C}_{2}=\frac{3}{%
16}$ in both cases. The Bargmann index $q$ allows us to deal with both cases
independently.

The $su(1,~1)$ Lie algebra operators satisfy
\begin{eqnarray}
K_{0}\left\vert q,n\right\rangle _{A} &=&(n+q)\left\vert q,n\right\rangle
_{A},  \notag \\
K_{+}\left\vert q,n\right\rangle _{A} &=&\sqrt{(n+q+\frac{3}{4})(n+q+\frac{1%
}{4})}\left\vert q,n+1\right\rangle _{A},  \notag \\
K_{-}\left\vert q,n\right\rangle _{A} &=&\sqrt{(n+q-\frac{3}{4})(n+q-\frac{1%
}{4})}\left\vert q,n-1\right\rangle _{A}.  \notag  \label{la}
\end{eqnarray}%
Projecting both sides of the Schr$\ddot{o}$dinger equation onto $\left\vert
q,n\right\rangle _{A}$ gives a linear relation between coefficients $%
e_{n}^{(q)}$ and $f_{n}^{(q)}$,
\begin{equation}
e_{n}^{(q)}=\frac{\Delta /2}{2\beta (n+q)-E+\frac{\epsilon -\omega }{2}}%
f_{n}^{(q)},  \label{en-r}
\end{equation}%
and a three-term linear recurrence relation is given by
\begin{widetext}
\begin{eqnarray}
f_{n+1}^{(q)}=\frac{2(2\omega ^{2}-\beta ^{2})(n+q)-\beta \left( E+\frac{%
\epsilon +\omega }{2}\right) -\frac{\Delta ^{2}\beta /4}{2\beta (n+q)
-E+\frac{\epsilon -\omega }{2}}}{8g\omega (n+q+\frac{1}{4})(n+q+\frac{3}{4})}%
f_{n}^{(q)}-\frac{1}{4(n+q+\frac{1}{4})(n+q+\frac{3}{4})}f_{n-1}^{(q)}.  \label{fn-r}
\end{eqnarray}
\end{widetext}
All coefficients $f_{n}^{(q)}$ and $e_{n}^{(q)}$ can be calculated with
initial conditions $f_{-1}^{(q)}=0$ and $f_{0}^{(q)}=1.$

We then apply the second squeezing operator $S_{B}=e^{-\frac{r}{2}%
(a^{2}-a^{\dag 2})}$ to the Hamiltonian (\ref{H_0}) and suggest the
wavefunction as%
\begin{equation}
\left\vert \Psi _{B}\right\rangle ^{q}=\binom{\sum_{m=0}^{\infty }(-1)^{m}%
\sqrt{[2(m+q-\frac{1}{4})]!}c_{m}^{(q)}\left\vert q,m\right\rangle _{B}}{%
\sum_{m=0}^{\infty }(-1)^{m}\sqrt{[2(m+q-\frac{1}{4})]!}d_{m}^{(q)}\left%
\vert q,m\right\rangle _{B}},  \label{w2}
\end{equation}%
where $\left\vert q,m\right\rangle _{B}=S_{B}\left\vert q,m\right\rangle $.
Similarly, we can obtain a linear relation between the coefficients $%
c_{n}^{(q)}$ and $d_{n}^{(q)}$
\begin{equation}
d_{n}^{(q)}=\frac{\Delta /2}{2\beta (n+q)-E-\frac{\epsilon +\omega }{2}}%
c_{n}^{(q)},  \label{dn-r}
\end{equation}%
and the three-term linear recurrence relation is
\begin{widetext}
\begin{eqnarray}
c_{n+1}^{(q)}=\frac{2(2\omega ^{2}-\beta ^{2})(n+q)-\beta (E-\frac{\epsilon
-\omega }{2})-\frac{\Delta ^{2}\beta /4}{2\beta (n+q)-E-\frac{\epsilon
+\omega }{2}}}{8g\omega (n+q+\frac{1}{4})(n+q+\frac{3}{4})}c_{n}^{(q)}-\frac{%
1}{4(n+q+\frac{1}{4})(n+q+\frac{3}{4})}c_{n-1}^{(q)}.  \label{cn-r}
\end{eqnarray}
\end{widetext}
Left-multiplying the vacuum state $\left\langle q,0\right\vert $ to the
extended squeezed state $\left\vert q,m\right\rangle _{A}$ and $\left\vert
q,m\right\rangle _{B}$, we can obtain the inner product
\begin{eqnarray}
\left\langle q,0\right\vert \left\vert q,m\right\rangle _{A} &=&\frac{%
(-\tanh r)^{m}}{\sqrt{\cosh r}}\frac{\sqrt{[2(m+q-\frac{1}{4})]!}}{2^{m}m!},
\notag \\
\left\langle q,0\right\vert \left\vert q,m\right\rangle _{B} &=&\frac{(\tanh
r)^{m}}{\sqrt{\cosh r}}\frac{\sqrt{[2(m+q-\frac{1}{4})]!}}{2^{m}m!}.
\end{eqnarray}

If both wavefunction $\left\vert \Psi _{A}\right\rangle ^{q}$ and $%
\left\vert \Psi _{B}\right\rangle ^{q}$ for the same $q$ are the true
eigenfunction for a non-degenerate eigenstate with eigenvalue $E$, they
should be proportional with each other, i.e. $\left\vert \Psi
_{A}\right\rangle ^{q}=z\left\vert \Psi _{B}\right\rangle ^{q}$, where $z$
is a complex constant. Projecting both sides of this identity onto the
original vacuum state $\left\vert q,0\right\rangle $, we obtain a
transcendental function below defined as G-function
\begin{eqnarray}
G^{(q)} &=&\left( \frac{\Delta }{2}\right) ^{2}\left[ \sum_{m=0}^{\infty }%
\frac{f_{m}^{(q)}\Omega _{m}^{\left( q\right) }}{2\beta (m+q)+\frac{\epsilon
-\omega }{2}-E}\right]  \notag \\
&&\times \left[ \sum_{m=0}^{\infty }\frac{c_{m}^{(q)}\Omega _{m}^{\left(
q\right) }}{2\beta (m+q)-\frac{\epsilon +\omega }{2}-E}\right]  \notag \\
&&-\sum_{m=0}^{\infty }f_{m}^{(q)}\Omega _{m}^{\left( q\right)
}\sum_{m=0}^{\infty }c_{m}^{(q)}\Omega _{m}^{\left( q\right) },  \label{G-tp}
\end{eqnarray}%
with%
\begin{equation*}
\Omega _{m}^{\left( q\right) }=\frac{(-\tanh r)^{m}}{\sqrt{\cosh r}}\frac{%
[2(m+q-\frac{1}{4})]!}{2^{m}m!}.
\end{equation*}%
If set $\epsilon =0$, the G-function for the symmetric tpQRM ~\cite{Chen2012}
is recovered. The zeros of the $G^{(q)}$-function give the regular spectrum
in the $q$ subspace of the asymmetric tpQRM.

From Eqs. (\ref{en-r}) and (\ref{dn-r}), we find the G-function diverges
when its denominators vanishes, the condition of the denominators being zero
can be obtained as
\begin{equation}
E_{m}^{A}=2\beta (m+q)+\frac{\epsilon -\omega }{2},  \label{pol1}
\end{equation}%
and%
\begin{equation}
E_{m}^{B}=2\beta (m+q)-\frac{\epsilon +\omega }{2},  \label{pol2}
\end{equation}%
with $m=0,1,2,...$. They are also labeled as two types (A and B) pole
energies, similar to the asymmetric QRM.

\begin{figure}[tbp]
\includegraphics[width=\linewidth]{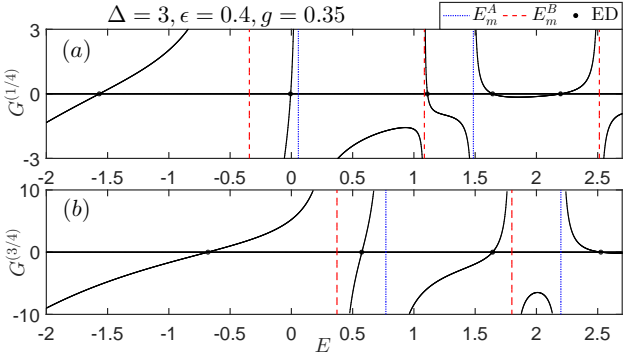}
\caption{ (Color online) $G$-curves for $\protect\omega=1, \Delta =3,\protect%
\epsilon =0.4,g=0.35$. $q=1/4$ (upper) and $q=3/4$ (lower). The blue dashed
lines are $E_{m}^{A}$ in Eq. (\protect\ref{pol1}) and red dashed lines are $%
E_{m}^{B}$ in Eq. (\protect\ref{pol2}), $m=0,1,2...$. The data by numerical
diagonalizations are indicated by black dots, which agree excellently with
the zeros of the G-functions (\protect\ref{G-tp}). }
\label{G-function}
\end{figure}

G-curves$\ $at $q=1/4$ and $3/4$ for $\epsilon =0.4,g=0.35,$ and $\Delta =3$
with $\omega =1$ are plotted in Fig. \ref{G-function}. The zeros are easily
detected. As usual, one can check it easily with numerics, an excellent
agreement can be achieved. The poles given in Eqs. (\ref{pol1}) and (\ref%
{pol2}) are marked with vertical lines. The G-curves indeed show diverging
behavior when approaching the poles.

In the limit of $g\rightarrow \omega /2$, the type-A pole energies are
squeezed into a single finite value $E_{m}^{A}(g=\omega /2)=\frac{\epsilon
-\omega }{2}$, and type-B pole energies into $E_{m}^{B}(g=\omega /2)=-\frac{%
\epsilon +\omega }{2}$. It seems that there are two kinds of collapse
energies $\frac{-\omega \pm \epsilon }{2}$. But actually, at $g=\omega /2$,
our obtained energy levels tend to the smaller one $\frac{-\omega
-\left\vert \epsilon \right\vert }{2}$, except some low lying states which
split off from the continuum. The ground-state is always separated from the
continuum by a finite excitation gap. In the spectrum shown in the next
sections, we will indeed observe that the plotted energy levels always
collapse to the smaller one at $g=\omega /2$, which generate more
non-degenerate exceptional points when $g\longrightarrow \omega /2$.
However, we cannot rule out the possibility that some energy levels would
stay between two limit energies $\frac{-\omega -\left\vert \epsilon
\right\vert }{2}$ and $\frac{-\omega +\left\vert \epsilon \right\vert }{2}$,
and thus these levels could not collapse. Since the analytical solution at $%
g=\omega /2$ in the asymmetric tpQRM is lacking, the collapse issue in this
model is somehow challenging.

\section{Non-degenerate exceptional solutions in the asymmetric tpQRM}

As outlined in the Sec. II (c) for the asymmetric one-photon QRM, we can
easily find the non-degenerate exceptional solutions in the spectra for the
asymmetric tpQRM by the pole structures of the G-function. When the energy
levels cross the pole lines, the coefficients in the G-function would
diverge, and therefore should be treated specially. For any real physical
systems, the wavefunction should be analytic, so the numerators $f_{m}^{(q)}$
in Eq. (\ref{en-r}) or $c_{m}^{\left( q\right) }$ Eq. (\ref{dn-r}) should
also vanish, which further gives the condition for the model parameters $%
g,\Delta ,\epsilon $, for fixed value of $m$ associated with one pole line.

In parallel to the asymmetric one-photon QRM, the first non-degenerate
exceptional G-function associated with the N-th type-A pole lines (\ref{pol1}%
) for the asymmetric tpQRM is easily given by
\begin{eqnarray}
G_{non,1A}^{(q)} &=&\left[ \sum_{n=0}^{N-1}\frac{\Delta f_{n}^{(q)}\Omega
_{n}^{\left( q\right) }/2}{2\beta \left( n-N\right) }+e_{N}\Omega
_{N}^{\left( q\right) }\right]  \notag \\
&&\times \left[ \sum_{n=0}^{\infty }\frac{\Delta c_{n}^{(q)}\Omega
_{n}^{\left( q\right) }/2}{2\beta \left( n-N\right) -\epsilon }\right]
\notag \\
&&-\sum_{n=0}^{N-1}f_{n}^{(q)}\Omega _{n}^{\left( q\right)
}\sum_{n=0}^{\infty }c_{n}^{(q)}\Omega _{n}^{\left( q\right) }=0,
\label{non_G1A}
\end{eqnarray}
where
\begin{equation}
e_{N}^{(q)}=-\frac{4g\omega }{\Delta \beta }f_{N-1}^{(q)},  \label{tp-eN}
\end{equation}%
and that associated with the M-th type-B pole lines (\ref{pol2}) reads
\begin{eqnarray}
G_{non,1B}^{(q)} &=&\left[ \sum_{n=0}^{\infty }\frac{\Delta
f_{n}^{(q)}\Omega _{n}^{\left( q\right) }/2}{2\beta \left( n-M\right)
+\epsilon }\right]  \notag \\
&&\times \left[ \sum_{n=0}^{M-1}\frac{\Delta c_{n}^{(q)}\Omega _{n}^{\left(
q\right) }/2}{2\beta \left( n-M\right) }+d_{M}\Omega _{M}^{\left( q\right) }%
\right]  \notag \\
&&-\sum_{n=0}^{\infty }f_{n}^{(q)}\Omega _{n}^{\left( q\right)
}\sum_{n=0}^{M-1}c_{n}^{(q)}\Omega _{n}^{\left( q\right) }=0,
\label{non_G1B}
\end{eqnarray}%
where
\begin{equation}
d_{M}^{(q)}=-\frac{4g\omega }{\Delta \beta }c_{M-1}^{(q)}.  \label{tp-dN}
\end{equation}%
Note that by Eq. (\ref{tp-eN}) [Eq. (\ref{tp-dN})], all the remaining
coefficients for $n>N$ [$n>M$] vanish. Zeros of the first non-degenerate
exceptional G-functions are equivalent to $f_{N}^{(q)}=0$ or $c_{M}^{(q)}=0$%
. Obviously, the later ones \ are obviously simpler in practical
calculations, while the former ones are more conceptually interesting, both
can give the same solutions.

Similarly, the second non-degenerate exceptional G-function associated with
the type-A pole lines (\ref{pol1}) is
\begin{eqnarray}
G_{non,2A}^{(q)} &=&\left[ \Omega _{N}^{\left( q\right)
}+\sum_{n=N+1}^{\infty }\frac{\Delta f_{n}^{(q)}\Omega _{n}^{\left( q\right)
}/2}{2\beta \left( n-N\right) }\right]  \notag \\
&&\times \left[ \sum_{n=0}^{\infty }\frac{\Delta c_{n}^{(q)}\Omega
_{n}^{\left( q\right) }/2}{2\beta \left( n-N\right) -\epsilon }\right]
\notag \\
&&-\sum_{n=N+1}^{\infty }f_{n}^{(q)}\Omega _{n}^{\left( q\right)
}\sum_{n=0}^{\infty }c_{n}^{(q)}\Omega _{n}^{\left( q\right) }=0,
\label{non_G1}
\end{eqnarray}%
where we have set $e_{N}^{(q)}=1$, and the coefficients $e_{n<N}^{(q)}=0$
and $f_{n\leqslant N}^{(q)}=0$. By the recurrence relations and the pole
energy, all other coefficients can be obtained. The second non-degenerate
exceptional G-function associated with the type-B pole lines (\ref{pol2})
can be obtained in a straightforward way as
\begin{eqnarray}
G_{non,2B}^{(q)} &=&\left[ \sum_{n=0}^{\infty }\frac{\Delta
f_{n}^{(q)}\Omega _{n}^{\left( q\right) }/2}{2\beta \left( n-M\right)
+\epsilon }\right]  \notag \\
&&\times \left[ \Omega _{M}^{\left( q\right) }+\sum_{n=M+1}^{\infty }\frac{%
\Delta c_{n}^{(q)}\Omega _{n}^{\left( q\right) }/2}{2\beta \left( n-M\right)
}\right]  \notag \\
&&-\sum_{n=0}^{\infty }f_{n}^{(q)}\Omega _{n}^{\left( q\right)
}\sum_{n=M+1}^{\infty }c_{n}^{(q)}\Omega _{n}^{\left( q\right) }=0,
\label{non_G2}
\end{eqnarray}%
where $d_{M}^{(q)}=1$, and the coefficients $d_{n<M}^{(q)}=0$ and $%
c_{n\leqslant M}^{(q)}=0$.

\begin{figure}[tbp]
\includegraphics[width=\linewidth]{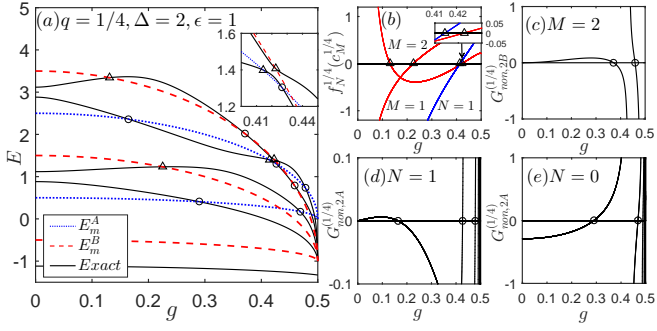}
\caption{ (Color online) (a) The spectra for the first $5$ levels and the
non-degenerate exceptional solutions for the asymmetric tpQRM with $\protect%
\omega=1, \Delta =2,\protect\epsilon =1,q=1/4$. The blue dashed lines are $%
E_{m=0,1}^{A}$ by Eq. (\protect\ref{pol1}) and red dashed lines are $%
E_{m=0,1,2}^{B}$ by Eq. (\protect\ref{pol2}). The inset on an enlarged scale
shows the avoided crossing instead of the true level crossing. $%
f_{N=1}^{(1/4)}$, $c_{M=1,2}^{(1/4)}$ curves are exhibited in (b), whose
zeros are indicated by open triangles, agreeing with the same symbols in the
spectra (a). The non-degenerate exceptional G-function $G_{non,2B}^{(1/4)} $
in Eq. (\protect\ref{non_G2}) for $M=2 $, and $G_{non,2A}^{(1/4)}$ in Eq. (%
\protect\ref{non_G1}) for $N=1,0$ are given in (c), (d), and (e),
respectively. Their zeros are denoted by open circles, which are excellently
consistent with the non-degenerate exceptional points indicated by the same
symbols in the spectra (a). }
\label{nondegenerate}
\end{figure}

We plot the spectra in Fig. \ref{nondegenerate} (a) for the parameters $%
q=1/4,\Delta =2,\epsilon =1$ with $\omega =1$. The crossing points of the
energy levels and the pole lines (\ref{pol1}) and (\ref{pol2}), known as
non-degenerate exceptional points, are marked with open symbols. All these
non-degenerate exceptional points can be confirmed analytically. The
solutions by the coefficient polynomial equations $f_{N=1}^{1/4}=0$ and $%
c_{M=1,2}^{1/4}=0$ are indicated by open triangles in Fig. \ref%
{nondegenerate} (b), and denoted with the same symbols in Fig. \ref%
{nondegenerate} (a). 7 zeros of the non-degenerate exceptional G-functions (%
\ref{non_G1}) and (\ref{non_G2}) corresponding to 7 open circles in Fig. \ref%
{nondegenerate} (c-e) are indicated by the 7 same symbols in Fig. \ref%
{nondegenerate} (a).

As revealed on an enlarged scale in the inset of Fig. \ref{nondegenerate}
(a) and (b) that two open triangles do not coincide, indicating an avoided
crossing at this bias parameter $\epsilon =1$. We will show in the left
panels of Fig. \ref{crossing1} at $\epsilon =1.0954$ in the next section
that, the two open triangles also obtained from $f_{N=1}^{(1/4)}=0$ and $%
c_{M=2}^{(1/4)}=0$ eventually can meet. Thus it should be very interesting
to see how an avoided crossing essentially turns to a true level crossing
when $\epsilon =1\rightarrow 1.0954$.

\section{Doubly degenerate states in asymmetric two-photon QRM}

In the asymmetric tpQRM, can we also find level crossings in the same $q$
subspace? According to the pole energies (\ref{pol1}) and (\ref{pol2}), if $%
E_{M}^{A}=E_{N}^{B}$, then%
\begin{equation}
\epsilon =2\beta \left( M-N\right) ,  \label{epsilon_p2}
\end{equation}%
the same pole energy takes
\begin{equation}
E=\left( M+N+2q\right) \beta -\frac{1}{2}\omega .  \label{pole_energy}
\end{equation}%
Interestingly, Eq. (\ref{epsilon_p2}) entails $\epsilon $ to be an even
multiple of the renormalized cavity frequency $\beta $, in contrast to the
asymmetric one-photon QRM where $\epsilon $ should be an multiple of the
cavity frequency $\omega $ under the condition (\ref{epsilon_p1}) for level
crossings. It makes sense that only the two-photon process is involved in
the two-photon model, while the single photon process in the one-photon
model.

Without loss of generality, we also only consider $M>N$ here. From Eq. (\ref%
{fn-r}) [(\ref{cn-r})], one immediately note that the coefficient $%
e_{N}^{(q)}$ in (\ref{en-r}) ($d_{M}^{(q)}$ in (\ref{dn-r})) would diverge
at the same pole energy (\ref{pole_energy}). Similar to the asymmetric QRM
case, the series expansion coefficients in the wavefunction (\ref{w1}) and (%
\ref{w2}) should be analytic and vanish as or before $n\rightarrow \infty $.

Regarding states with the energy (\ref{pole_energy}), the numerator of
right-hand-side of (\ref{fn-r}) [(\ref{dn-r})] should also vanish, so that $%
e_{N}^{(q)}$ ($d_{M}^{(q)}$) remains finite, which requires%
\begin{equation}
f_{N}^{(q)}(M,g)=0;c_{M}^{(q)}(N,g)=0.  \label{ocation_2p}
\end{equation}%
Note that $f_{N}^{(q)}$ and $c_{M}^{(q)}$ can be respectively obtained from
the recurrence relations (\ref{fn-r}) and (\ref{cn-r}) by using the same
pole energy (\ref{pole_energy})
\begin{widetext}
\begin{eqnarray}
f_{n+1}^{(q)}=\frac{2\omega ^{2}\left( n+q\right) -\beta ^{2}\left(
n+M+2q\right) +\frac{\Delta ^{2}}{16\left( N-n\right) }}{4g\omega (n+q+%
\frac{1}{4})(n+q+\frac{3}{4})}f_{n}^{(q)}-\frac{1}{4(n+q+\frac{1}{4})(n+q+%
\frac{3}{4})}f_{n-1}^{(q)}, \label{2pfrn}
\end{eqnarray}
\begin{eqnarray}
c_{n+1}^{(q)}=\frac{2\omega ^{2}\left( n+q\right)-\beta ^{2}\left(
n+N+2q\right) +\frac{\Delta ^{2}}{16\left( M-n\right) }}{4g\omega (n+q+%
\frac{1}{4})(n+q+\frac{3}{4})}c_{n}^{(q)}-\frac{1}{4(n+q+\frac{1}{4})(n+q+%
\frac{3}{4})}c_{n-1}^{(q)}. \label{2pcrn}
\end{eqnarray}
\end{widetext}

Similar to the asymmetric one-photon QRM, we conjecture \ that both $%
f_{N}^{(q)}(M,g)=0$ and $c_{M}^{(q)}(N,g)=0$ could give the same positive
real $g$ and $\epsilon $ under the constrained condition (\ref{epsilon_p2}),
leading to levels crossing at the same pole energy. While it would be
interesting to rigorously prove the conjecture in the two-photon case
mathematically, we also confine us here to an analytical closed-form proof
only for small values of $N$ and $M$, and numerically confirmation for large
$N$ and $M$, in searching for physically reasonable coupling strength $g$.
Similar to the asymmetric QRM, we also present our discussions only in terms
of fixed values of $N$ and $M,$ but here $\epsilon $ cannot be determined
independently, and would be determined together with $g$ by Eqs. (\ref%
{epsilon_p2}) and (\ref{ocation_2p}).

{In Appendix A3,} we analytically prove that, for some small values of $N$
and $M$, both $f_{N}^{(q)}(M,g)=0$ and $c_{M}^{(q)}(N,g)=0$ in (\ref%
{ocation_2p}) give the same values for $\epsilon $ and $g$. Two energy
levels cross the corresponding pole lines at the same values of $\epsilon $
and $g$, where the two pole lines also cross. Thus true level crossings also
happen in the asymmetric tpQRM. Compare to the one-photon QRM where $%
\epsilon $ can be determined independently, in the asymmetric tpQRM, we need
to solve two equations simultaneously to determine $\epsilon $ and $g$.

\begin{figure}[tbp]
\includegraphics[width=\linewidth]{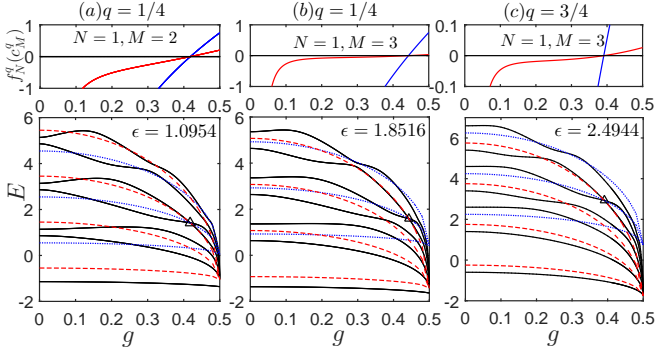}
\caption{ (Color online) Energy spectrum are list in low panels. The black
lines are energy levels, the blue dashed lines are $E_{m}^{A}\ $ and the red
dashed lines are $E_{m}^{B}$. Open triangles indicate the doubly degenerate
level crossings. $f_{N=1}^{(q)}(M,g)\ $(blue) and $c_{M}^{(q)}(N=1,g)$ (red)
curves are displayed in the upper panels. $\left( q,\protect\epsilon \right)
$ = $\left( 1/4,1.0954\right) $ (left), $\left( 1/4,1.8516\right) $
(middle), and $\left( 3/4,2.4944\right) $ (right). Zeros are the same for
both curves. $\Delta =2$ and $\protect\omega=1$. }
\label{crossing1}
\end{figure}

We show the energy spectrum of the asymmetric tpQRM at $N=1$, $\Delta =2$
with $\omega =1$, for $q=1/4,M=2$ (left), $q=1/4,M=3$ (middle), and $%
q=3/4,M=3$ (right) in the low panels of Fig. \ref{crossing1}. The
corresponding values of $\epsilon $ are just those determined by Eq. (\ref%
{epsilon_m}), which in turn are $\epsilon =1.0954,1.8516$ , and $\allowbreak
2.\,\allowbreak 4944$ from left to right. Interestingly, one level crossing
point indicated by the open triangle really appears in each spectra,
confirming the analytical prediction.

The upper panels in Fig. \ref{crossing1} present the curves for $%
f_{1}^{(q)}(M,g)$ and $c_{M}^{(q)}(1,g)$. It is clear that the zeros of both
functions are the same, and are consistent with the coupling strength at the
level crossing points. For example, for $q=1/4$, $\Delta =2$, two energy
levels cross exactly at $g=0.4183$ by Eq. (\ref{gN1M2}). This analytical
findings is in excellent consistent with numerical results presented in the
left panels of Fig. \ref{crossing1}. This agreements also applies to the
middle and right panels. As expected, the type-A pole lines $E_{N=1}^{A}$
and the type-B pole lines $E_{M}^{B}$ also cross at the degenerate points in
the low panels of Fig. \ref{crossing1}.

\begin{figure}[tbp]
\includegraphics[width=\linewidth]{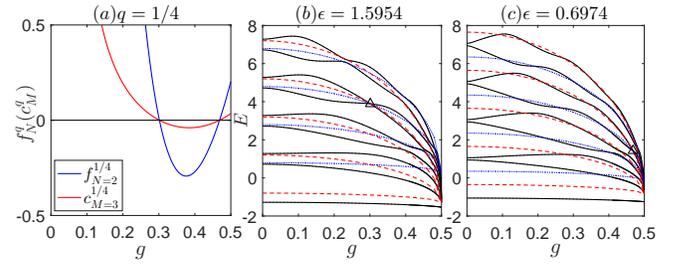}
\caption{ (Color online) $f_{N=2}^{(1/4)}$ and $c_{M=3}^{(1/4)}$ curves are
exhibited in (a). Zeros are the same for both curves. Energy spectrum are
presented in (b) for $\protect\epsilon =1.5954$ and (c) $\protect\epsilon %
=0.6974$. The black lines are energy levels, the blue dashed lines are $%
E_{m}^{A},$ and the red dashed lines are $E_{m}^{B}$. Open triangles
indicate the doubly degenerate level crossings. $\Delta =2$ and $\protect%
\omega=1$. }
\label{crossing12}
\end{figure}

For $N=1$, no matter what is the value of $M>N$, from Eq. (\ref{closed}), we
can at most find one solution for $g,$ which is $M$ dependent. For $N>1$, $%
f_{N}^{(q)}(M,g)=0$ is a polynomial equation with $N$ terms, which would
give more than one solutions for $g$, and further corresponding solutions
for $\epsilon $ in terms of Eq. (\ref{epsilon_p2}).

As shown in left panel of Fig. \ref{crossing12} for $\Delta =2, \omega=1,
q=1/4, N=2$ and $M=3$ , both $f_{2}^{(1/4)}(3,g)=0$ and $%
c_{3}^{(1/4)}(2,g)=0 $ yield the same solutions for $g_{1,2}=0.3015,0.4686$ {%
\ by Eq. (\ref{g1N2M3}) , and two values of $\epsilon _{1,2}=1.5954,0.6974$}
{\ are then determined accordingly. We then plot the energy spectrum for
these two values of $\epsilon $ in the middle and right panels of Fig. \ref%
{crossing12} for $\Delta=2$. The level crossings are clearly shown at the
analytical predicted coupling strength. Note that the $N=2$ type-A pole line
and the $M=3$ type-B pole line indeed cross at the same doubly degenerate
points. }

Finally, the doubly degenerate states at the {true level crossing points can
be expressed explicitly in terms of }the BOA as
\begin{equation*}
\left\vert \Psi _{A}\right\rangle ^{q}=\binom{\sum_{m=0}^{N}\sqrt{[2(m+q-%
\frac{1}{4})]!}e_{m}^{(q)}\left\vert q,m\right\rangle _{A}}{\sum_{m=0}^{N-1}%
\sqrt{[2(m+q-\frac{1}{4})]!}f_{m}^{(q)}\left\vert q,m\right\rangle _{A}},
\end{equation*}
and
\begin{equation*}
\left\vert \Psi _{B}\right\rangle ^{q}=\binom{\sum_{m=0}^{M-1}(-1)^{m}\sqrt{%
[2(m+q-\frac{1}{4})]!}c_{m}^{(q)}\left\vert q,m\right\rangle _{B}}{%
\sum_{m=0}^{M}(-1)^{m}\sqrt{[2(m+q-\frac{1}{4})]!}d_{m}^{(q)}\left\vert
q,m\right\rangle _{B}},
\end{equation*}%
respectively, where $e_{N}^{(q)}$ and $d_{M}^{(q)}$ are given by Eqs (\ref%
{tp-eN}) and (\ref{tp-dN}). Because these two wavefuntions are not obtained
from the G-function based on the proportionality, so they are different,
leading to {doubly degenerate states}. Both wavefunction terminates at
finite terms, so they are the quasi-exact solutions of the asymmetric tpQRM.

\section{Discussions}

\begin{figure}[tbp]
\includegraphics[width=\linewidth]{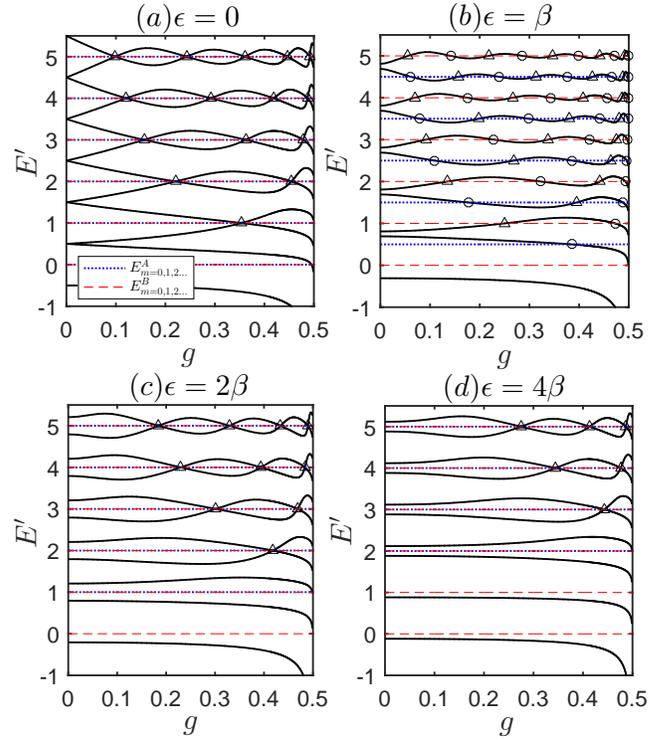}
\caption{ Energy spectrum $E^{\prime }=\frac{E+1/2}{2\protect\beta }-q+\frac{
\protect\epsilon }{4\protect\beta }$ for $\protect\omega=1, \Delta =2,q=1/4,
\protect\epsilon = k \protect\beta $. $k=0$ (a), $1$ (b), $2$ (c), and $4$
(d). Note particularly that $\protect\epsilon$ changes with $g$ along $g$%
-axis. The horizontal blue dotted lines correspond to the pole energy ones
for $E_{N}^{A}$ and the red dashed lines to $E_{M}^{B}$. Only the overlapped
pole lines with $N>0$ allow for the true level crossings. The triangles
denotes the doubly degenerate crossing points in (a), (c), and (d). In (b),
the triangles are obtained from $f_{N}^{(q)}=0$ and $c_{M}^{(q)}=0$ or
equivalently from Eqs. ( \protect\ref{non_G1A}) and (\protect\ref{non_G1B}),
while open circles from Eqs. ( \protect\ref{non_G1}) and (\protect\ref%
{non_G2}), all of them correspond to non-degenerate exceptional points. }
\label{spectr_tp}
\end{figure}

From the spectrum in Figs. \ref{crossing1} and \ref{crossing12}, one might
speculate that level crossings seldom happen in the asymmetric tpQRM.
Actually it is not that case. If we incorporate Eq. (\ref{epsilon_p2})
required by the level crossings, we may plot the similar spectra graph as
Figs. \ref{1pcrossD15} and \ref{1pcrossD30} in one-photon case. In doing so,
we calculate the energy as a function of $g$, and at the same time $\epsilon
$ also changes as {Eq.} (\ref{epsilon_p2}). To display the level crossings
in asymmetric tpQRM more clearly, we can make the pole lines horizontal,
thus we plot the normalized energy $E^{\prime }=\frac{E+\omega /2}{2\beta }%
-q+\frac{\epsilon }{4\beta }$ as \ a function of $g$ and simultaneously
varying $\epsilon =k\beta $ in Fig. \ref{spectr_tp} for $k=0,1,2,4$ at $%
\Delta =2$ with $\omega =1,q=1/4$.

When $\epsilon $ is an even multiple of the normalized cavity frequency $%
\beta $ entailed in Eq. (\ref{epsilon_p2}), i.e. $k$ is an even integer
including the symmetric case $k=0$, we find that the two equations in Eq. (%
\ref{ocation_2p}) result in the same positive solutions for the coupling
strength, as indicated with open triangles (a), (c) and (d). One can note
that the level crossings happen regularly. The crossing points at the $N=1$
type-A pole line in (c) and (d) are just corresponding to those in Figs. \ref%
{crossing1} (a) and (b), while the two crossing points at the $N=2$ type-A
pole line in (c) to those in Fig. \ref{crossing12}.

However, if $k$ is not an even integer, no level crossings happen,  a lot of
non-degenerate exceptional points emerges instead. As exhibited in Fig. \ref{spectr_tp} (b) for $
k=1$, the open triangles correspond to  the non-degenerate exceptional points
by Eqs. (\ref{non_G1A}) or (\ref{non_G1B}), while the open circles to those by
 Eqs. (\ref{non_G1}) and (\ref{non_G2}).

{We can also estimate the number of the doubly degenerate crossing points
associated with the given N type-A pole line. } For any $M>N$, generally
there are around $N$ crossing points due to the polynomial equation with $N$
terms, $f_{N}^{(q)}(M,g)=0$ in Eq. (\ref{ocation_2p}), the detailed
polynomial equations is derived from Eq. (\ref{2pfrn}). This is to say,
associated with N type-A pole line, we generally have around $N$ degenerate
crossing points for both asymmetric one-photon and two-photon QRMs. In
Appendix A3, we have employed the constrained conditions in the asymmetric
both one- and two-photon QRMs, and numerically found that they have nearly
the same numbers of level crossing points in the range of integers of $N$
and $M$ in each case. Therefore, we could reach a conclusion that the number
of the doubly degenerate crossing points in asymmetric tpQRM would be twice
of that in the asymmetric QRM due to two Bargmann indices in the former
model.

Braak proposed a new criterion of integrability that if the eigenstates of a
quantum system can be uniquely labeled by $i=i_{1}+i_{2}$ quantum numbers,
where $i_{1}$ and $i_{2}$ are the numbers of the discrete and continuous
degree of freedom, then it is integrable \cite{Braak}. Both symmetric QRM
and tpQRM may be considered integrable in terms of this criterion. \ As the
bias term of qubit sets in, the integrability will be violated in the
asymmetric QRM. However, if $\epsilon $ matches the multiple of the cavity
frequency, the integrability can be recovered in the asymmetric QRM, by
using the hidden symmetry instead of the parity number. As shown in Figs. %
\ref{1pcrossD15} and \ref{1pcrossD30}, the regular level crossings reappear
when $\epsilon/\omega $ is an integer, similar to that in the symmetric QRM
which is considered to be integrable ~\cite{Braak}. However, the asymmetric
tpQRM with fixed $\epsilon $ is always non-integrable because the energy
levels cannot be uniquely labeled by the only continuous degree of freedom.
As displayed in the spectrum in Figs. \ref{crossing1} and \ref{crossing12}
with special $\epsilon $'s, there is no regular level crossings, in sharp
contrast to the integrable symmetric tpQRM~\cite{duan2016}. Of course, if $%
\epsilon $ changes as $k\beta $ with $k$ an even integer, the regular level
crossings reappear in the asymmetric tpQRM as shown in Fig. \ref{spectr_tp},
and it can be reconsidered to be integrable.

In the asymmetric QRM, the effort to look for the hidden symmetry
responsible for the level crossings in the same $\epsilon $, continues to be
a great interest \cite{lizimin1, Batchelor,lizimin2,Wakayama}. Since the
doubly degenerate states within the same $q$ subspace also exist in the
asymmetric tpQRM, which is definitely not owing to an explicit symmetry. It
should be also interesting to rigorously find hidden symmetry in the
asymmetric tpQRM in the near future.

\section{Conclusion}

In this paper, we have studied both the asymmetric QRM and the asymmetric
tpQRM by the BOA in a unified way. The previously observed level crossing
when the bias parameter $\epsilon $ is a multiple of cavity frequency in the
asymmetric QRM is illustrated by a closed-from proof for low orders of the
constrained polynomial equations in a transparent manner. For the asymmetric
tpQRM, the biased term breaks original $\mathbb{Z}_{4}$ symmetry to $\mathbb{%
Z}_{2}$ symmetry, so the Hilbert space only divides into even and odd
bosonic number state subspaces. In each subspace, we derived the
transcendental equation, called G-function, and obtain the regular spectrum
exactly. The coefficients at the pole energy vanish in two different ways,
giving two kinds of non-degenerate exceptional G-functions, by which all
non-degenerate exceptional points can be detected.

Very interestingly, the true level crossings can also happen in the same $q$
subspace in the asymmetric tpQRM if the qubit bias parameter $\epsilon $ is
an even multiple of the $g$-dependent renormalized cavity frequency, in
contrast to the asymmetric one-photon QRM where $\epsilon $ can be simply a
multiple of the cavity frequency. We argue that the even multiple is
originated from the two-photon process involved in the two-photon model. The
doubly degenerate points can be also located analytically, similar to the
asymmetric QRM. The number of the doubly degenerate points within the same
subspace in the asymmetric tpQRM should be comparable with that in
asymmetric QRM. The subspace in the asymmetric tpQRM has no any explicit
symmetry, the newly found double degeneracy thus also implies the hidden
symmetry. The hidden symmetry in the asymmetric QRM could be identified at
the same integer $\epsilon /\omega $, while in the asymmetric tpQRM at the
same integer $\epsilon /(2\beta )$. The latter constraint on the parameter
space for the occurrence of the double degeneracy is illuminating in
searching for a conserved operator in two-photon case. The present results
may shed some lights on the different nature of the hidden symmetries in the
two asymmetric QRMs.

\textbf{ACKNOWLEDGEMENTS}  This work is supported by the
National Science Foundation of China under No. 11834005, the National Key
Research and Development Program of China under No. 2017YFA0303002.

$^{\ast }$ Email:qhchen@zju.edu.cn

\begin{appendix}

\section{Demonstration for the same physical solutions of the two equations
in the constrained conditions in two asymmetric QRMs}

In this Appendix, we first present a closed-form proof for the conjecture
that $f_{N}(M,g)=0$ and $c_{M}(N,g)=0$ in Eq. (\ref{location}) could give
the same real and positive solutions for the coupling strength $g$ with
small numbers of $N$ and $M$ in the asymmetric one-photon QRM. In parallel,
we then provide a closed-form proof for the conjecture that $%
f_{N}^{(q)}(M,g)=0$ and $c_{M}^{(q)}(N,g)=0 $ in Eq. (\ref{ocation_2p})
could give the same real and positive solutions for the coupling strength $g$
with small numbers of $N$ and $M$ in the asymmetric tpQRM. Finally, we
provide numerical confirmations on the conjecture with large range of
integers $N$ and $M$ in two asymmetric QRMs.  We set  $\omega=1$ in both models for simplicity in the whole Appendix.

\subsection{Analytical proof for the small order of the constrained conditions in asymmetric one-photon QRM}

Since $f_{0}=1$, we begin with the $N=1\ $type-A pole energy, Eq. (\ref{r_fN}%
) becomes%
\begin{equation*}
f_{1}{\left( M,g\right) }=\frac{1}{2g}\left( 4g^{2}+\frac{1}{4}\Delta
^{2}-M\right) ,
\end{equation*}%
its zero is simply
\begin{equation}
g=\frac{1}{2}\sqrt{M-\left( \frac{\Delta }{2}\right) ^{2}},  \label{gN1}
\end{equation}%
which is dependent on $M$. If $\Delta >2\sqrt{M}$, no real solution exists,
so the level crossing dose not occur along the $N=1$ pole line.

If we set $M=2$ i.e. $\epsilon =1$, we have
\begin{equation}
g=\frac{1}{2}\sqrt{2-\left( \frac{\Delta }{2}\right) ^{2}}.  \label{1pN1M2}
\end{equation}%
The second equation in (\ref{location})$\ {c_{2}\left( 1,g\right) }\bigskip
=0$ yields%
\begin{equation*}
\left( 4g^{2}+\frac{\Delta ^{2}}{4}\right) \left( 4g^{2}+\frac{\Delta ^{2}}{8%
}-1\right) -4g^{2}=0,
\end{equation*}%
resulting in%
\begin{equation*}
g=\frac{1}{2}\sqrt{2-\left( \frac{\Delta }{2}\right) ^{2}},
\end{equation*}%
which is {exactly the same as Eq. (\ref{1pN1M2}), the solution for $%
f_{1}\left( 2,g\right) =0$. It follows that two energy levels intersect with
the same pole line at the same coupling strength }$g${\ in the spectra,
indicating a true energy level crossing. }

For \ $N=2\ $type-A pole energy, the first equation in (\ref{location}) {$%
f_{2}\left( M,g\right) =0$ } becomes ({\ we set }${x=4g}^{2}$  for
simplicity)%
\begin{equation}
\left( x+\frac{1}{4}\Delta ^{2}-M+1\right) \left( x+\frac{1}{8}\Delta
^{2}-M\right) -x=0,  \label{N2}
\end{equation}%
yielding%
\begin{equation*}
x=\left( M-\frac{3}{16}\Delta ^{2}\right) \pm \sqrt{\left( \frac{\Delta ^{2}%
}{16}-1\right) ^{2}+\left( M-1\right) }.
\end{equation*}%
If $M=3$, i. e. $\epsilon $ is still $1$, the solutions then read%
\begin{equation}
g=\frac{1}{8}\sqrt{-3\left( \Delta ^{2}-16\right) \pm \sqrt{\left( \Delta
^{2}-16\right) ^{2}+512}}.  \label{N2M3}
\end{equation}%
On the other hand, the second equation in (\ref{location}) $c{_{3}\left(
2,g\right) }${$=0\ $is }
\begin{eqnarray}
&&\left( x+\frac{\Delta ^{2}}{4}\right) \left[ \left( x+\frac{\Delta ^{2}}{8}%
-1\right) \left( x+\frac{\Delta ^{2}}{12}-2\right) -x\right]  \notag \\
&&-2x\left( x+\frac{\Delta ^{2}}{12}-2\right) =0,  \label{NM3}
\end{eqnarray}%
which interestingly gives the same solutions as in Eq. (\ref{N2M3}){,
consistent with the conjecture}. Here an unphysical solution $x=-\frac{1}{12}%
\Delta ^{2}$ is omitted.

Next, we set $N=1,M=3$, thus $\epsilon =2$. $f_{1}{\left( 3,g\right) }=0$
gives%
\begin{equation}
g=\frac{1}{4}\sqrt{12-\Delta ^{2}}.  \label{N1M3}
\end{equation}%
By $c_{3}(1,g)$ $=0$, we have
\begin{eqnarray*}
&&\left( x+\frac{1}{4}\Delta ^{2}+1\right) \left[ \left( x+\frac{1}{8}\Delta
^{2}\right) \left( x+\frac{1}{12}\Delta ^{2}-1\right) -x\right] \\
&&-2x\left( x+\frac{1}{12}\Delta ^{2}-1\right) =0.
\end{eqnarray*}%
Its solutions are
\begin{equation*}
x=3-\frac{1}{4}\Delta ^{2};x=-\frac{\Delta ^{2}}{48}\left( 5\pm \sqrt{1-%
\frac{96}{\Delta ^{2}}}\right) .
\end{equation*}%
Note that the second root is not a positive real value, and so omitted. The
first root gives exactly the same $g$ in Eq. (\ref{N1M3}).

\subsection{Analytical proof for the small order of the constrained conditions in asymmetric tpQRM}

In this Appendix, we present a closed-form proof for the conjecture that $%
f_{N}^{(q)}(M,g)=0$ and $c_{M}^{(q)}(N,g)=0 $ in Eq. (\ref{ocation_2p})
could give the same real and positive solutions for the coupling strength $g$
with small numbers of $N$ and $M$ in the asymmetric tpQRM.

For the most simply case, we set $N=1$, then $f_{1}^{(q)}(M,g)=0$ gives
\begin{equation}
4q-\left( 2M+4q\right) \beta ^{2}+\frac{\Delta ^{2}}{8}=0,  \label{closed}
\end{equation}
then the location of the degenerate point is obtained
\begin{equation}
\beta ^{2}=\frac{2q+\Delta ^{2}/16}{M+2q},  \label{gM}
\end{equation}%
$\allowbreak $which is dependent on $M$. Also note that the positive real
solution only exists for $\Delta <4\sqrt{M}$. Subject to the constrained
condition (\ref{epsilon_p2}), we have%
\begin{equation}
\epsilon =\left( M-1\right) \sqrt{\frac{8q+\Delta ^{2}/4}{2q+M}}.
\label{epsilon_m}
\end{equation}%
If set {$M=2,{c_{2}^{(q)}}(1,g)=0$ gives}%
\begin{eqnarray*}
&&\left[ 4\left( q+1\right) \left( 1-\beta ^{2}\right) +\frac{1}{8}\Delta
^{2}\right]  \notag \\
&&\times \left[ 2\left( 2q+1\right) \left( 1-\beta ^{2}\right) +\frac{1}{16}\Delta
^{2}-2\right] \\
&&-4(q+\frac{1}{4})(q+\frac{3}{4})\left( 1-\beta ^{2}\right) =0,
\end{eqnarray*}%
we then have
\begin{equation}
\beta ^{2}=\frac{2q+\Delta ^{2}/16}{2+2q},  \label{gN1M2}
\end{equation}%
which is the same as that in Eq. (\ref{gM}) for $M=2${, consistent with our
conjecture.}

Next, we set $N=2,M=3.$ $f_{2}^{(q)}(3,g)=0$ gives
\begin{eqnarray*}
&&\left[ 4\left( 1-\beta ^{2}\right) \left( 2+q\right) +\frac{\Delta ^{2}}{8}%
-4\right]  \notag \\
&&\times \left[ 2\left( 1-\beta ^{2}\right) \left( 3+2q\right) +\frac{\Delta ^{2}}{%
16}-6\right] \\
&&-4\left( 1-\beta ^{2}\right) (q+\frac{1}{4})(q+\frac{3}{4})=0.
\end{eqnarray*}%
The solutions at $q=\frac{1}{4}$ are
\begin{equation*}
\beta ^{2}=\frac{1}{3}+\frac{23\Delta ^{2}}{2016}\pm \frac{1}{126}\sqrt{%
\frac{25\Delta ^{4}}{256}+\frac{21}{2}\Delta ^{2}+1008},
\end{equation*}
and at $q=\frac{3}{4}$ are
\begin{equation*}
\beta ^{2}=\frac{5}{11}+\frac{29\Delta ^{2}}{3168}\pm \sqrt{\frac{20}{363}+%
\frac{\Delta ^{2}}{8712}+\frac{49\Delta ^{4}}{3168^{2}}},
\end{equation*}%
while $c_{3}^{(q)}(2,g)=0$ results in
\begin{eqnarray*}
&&\{\left( 2+q+\frac{\Delta ^{2}}{32\left( 1-\beta ^{2}\right) }\right) %
\left[ 2\left( 1-\beta ^{2}\right) \left( 3+2q\right) +\frac{\Delta ^{2}}{16}%
-2\right] \\
&&-(q+\frac{5}{4})(q+\frac{7}{4})\}\times \left[ 4\left( 1-\beta ^{2}\right) \left(
1+q\right) +\frac{\Delta ^{2}}{24}-4\right] \\
&&-\left[ 4\left( 1-\beta ^{2}\right) \left( 2+q\right) +\frac{\Delta ^{2}}{8%
}\right] (q+\frac{1}{4})(q+\frac{3}{4})=0.
\end{eqnarray*}

If $q=\frac{1}{4}$, the solutions are
\begin{eqnarray*}
\beta ^{2} &=&\frac{1}{3}+\frac{23\Delta ^{2}}{2016}\pm \frac{1}{126}\sqrt{%
\frac{25\Delta ^{4}}{256}+\frac{21}{2}\Delta ^{2}+1008}, \\
\beta _{3}^{2} &=&1+\frac{\Delta ^{2}}{120}.
\end{eqnarray*}%
If $q=\frac{3}{4}$, the solutions are
\begin{eqnarray*}
\beta ^{2} &=&\frac{5}{11}+\frac{29\Delta ^{2}}{3168}\pm \sqrt{\frac{20}{363}%
+\frac{\Delta ^{2}}{8712}+\frac{49\Delta ^{4}}{3168^{2}}}, \\
\beta _{3}^{2} &=&1+\frac{\Delta ^{2}}{168}.
\end{eqnarray*}%
Omitting the unreasonable solutions $\beta _{3}$, we can find that both $%
f_{2}^{(q)}(3,g)=0$ and $c_{3}^{(q)}(2,g)$ give the same crossing coupling
strengths for $q=1/4$ and $3/4$ respectively {\
\begin{equation}
g_{1,2}^{(1/4)}=\frac{1}{2}\sqrt{\frac{2}{3}-\allowbreak \frac{23}{2016}%
\Delta ^{2}\pm \frac{\sqrt{25\Delta ^{4}+2688\Delta ^{2}+258048}}{2016}},
\label{g1N2M3}
\end{equation}%
}

\begin{equation}
g_{1,2}^{(3/4)}=\frac{1}{2}\sqrt{\frac{6}{11}-\frac{29}{3168}\allowbreak
\Delta ^{2}\pm \frac{\sqrt{49\Delta ^{4}+1152\Delta ^{2}+552960}}{3168}}.
\label{g3N2M3}
\end{equation}%
which also agree well with our conjecture.

\subsection{Numerical confirmation for the two conjectures in both
asymmetric QRMs}

\begin{figure}[tbp]
\includegraphics[width=\linewidth]{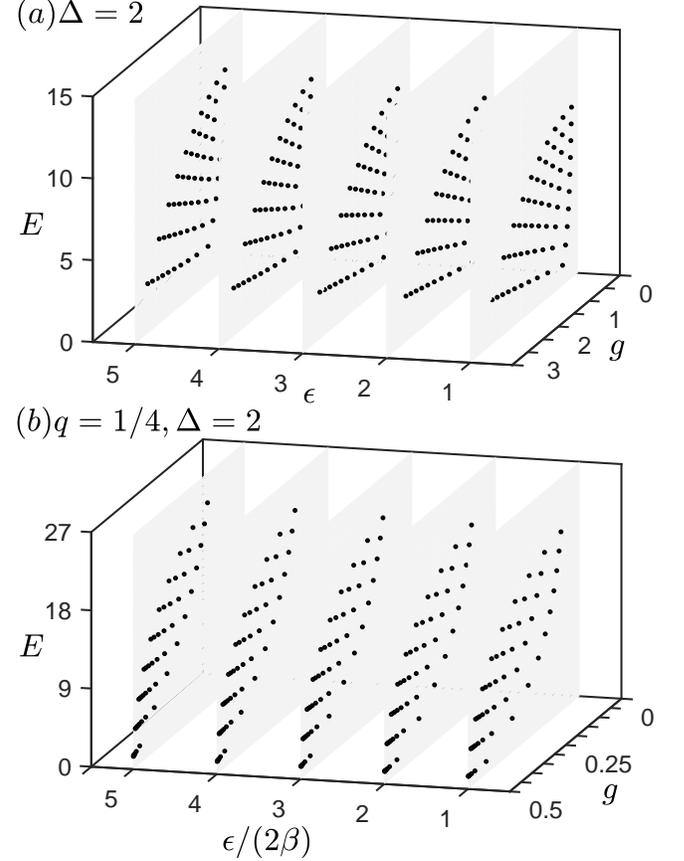}
\caption{ Three-dimensional view for the doubly degenerate level crossing
points at $\omega=1, \Delta =2$ for the asymmetric one-photon QRM (a) and tpQRM in the
$q=1/4$ subspace (b). We set $N$ from $1$ to $10$ and $M$ from $2$ to $20$
for both cases, and the numbers of the true level crossing points are the
same. The data are drawn from part of the level crossings in this range. }
\label{3Dview}
\end{figure}

We extensively demonstrate that, for large $N$ and $M(>N)$, the two equations  in
either Eq. (\ref{location}) or Eq. (\ref{ocation_2p}) give
the same physics solutions in both asymmetric QRMs. We sets $N$ from $1$ to $%
10$ and $M$ from $2$ to $20$ for both one-photon QRM and tpQRM in the $q=1/4$
subspace at $\Delta =2$. First, we  find that physics solutions from $f_{N}=0
$ and $c_{M}=0$ are exactly the same in either case, confirming the
conjectures numerically. Second, there are $715$ level crossings points for
both cases, indicating $N$ roots in the $N$ order polynomial equations in both models at $%
\Delta =2$. Generally, the root number is equal to or slightly less than $N$
for any $\Delta $. This is to say, for any values of $\Delta $, the numbers
of the level crossings are generally nearly the same for the same ranges of $%
N$ and $M$ in asymmetric QRMs.

In Fig. \ref{3Dview}, the doubly degenerate level crossing points are
visualized in a three-dimensional (3D) view in ($\epsilon , g, E$)-space for the
asymmetric one-photon QRM and in ($\epsilon /2\beta, g, E$)-space for the asymmetric tpQRM at $\Delta =2$. It is interesting to draw
planes for level crossings in both cases, as $\epsilon$ is simply scaled by
a $g$-dependent factor, $2\beta $, in the two-photon case. In the original 3D ($%
\epsilon, g, E$)-space, all the degenerate crossing points in asymmetric
QRM are confined in equally spaced integer $\epsilon/\omega  $ planes, while those
in asymmetric tpQRM are actually locked in different cylindrical surfaces
with integer $\epsilon /\left( 2\beta \right) $. Those different constrained
surfaces in the model parameter spaces for the occurrence of the double
degeneracy in two models should be considered in the definition of conserved operators and the detection of
hidden symmetries.

\end{appendix}


\end{document}